\documentclass[aps,prb,reprint,superscriptaddress,showpacs,floatfix]{revtex4-1}
\usepackage{graphicx}% Include figure files
\usepackage{amssymb,amsmath}
\usepackage{dcolumn}% Align table columns on decimal point
\usepackage{bm}% bold math
\usepackage[mathlines]{lineno}% Enable numbering of text and display math
\begin{document}

\title{Oxidized silicon sulfide: stability and electronic properties of
       a novel two-dimensional material}

\author{Zhengnan Li}
\affiliation{School of Physics,
             Southeast University,
             Nanjing 211189, PRC}

\author{Shuai Dong}
\affiliation{School of Physics,
             Southeast University,
             Nanjing 211189, PRC}

\author{Jie Guan}
\email%[E-mail: ]%
            {guanjie@seu.edu.cn}%
\affiliation{School of Physics,
             Southeast University,
             Nanjing 211189, PRC}

%\date{\today} % delete this line to display the current date

%---------------------------------------------------------------------
\begin{abstract}
Isolated oxygen impurities and fully oxidized structures of four stable two-dimensional (2D) SiS structures are investigated by {\em ab initio} density functional calculations. Binding energies of oxygen impurities for all the four 2D SiS structures are found larger than those for phosphorene, due to the lower electronegativity of Si atoms. The most stable configurations of isolated oxygen impurities for different 2D SiS structures are decided and the corresponding 2D structures with saturated oxidation (SiSO) are predicted. Among all the four fully oxidized structures, $\alpha$-SiSO is demonstrated to be stable by phonon spectra calculations and molecular dynamics (MD) simulations. Electronic structure calculations indicate that $\alpha$-SiSO monolayer is semiconducting with a direct band gap of ${\approx}2.28$~eV, which can be effectively tuned by in-layer strain. The value of band gap and thermodynamic stability are found depending sensitively on the saturation level of oxygen.
\end{abstract}
%---------------------------------------------------------------------


\maketitle
%---------------------------------------------------------------------
% \linenumbers\relax % Commence numbering lines
% If in two-column mode, this environment will change to single-column
% format so that long equations can be displayed. Use sparingly.
%\begin{widetext}
%\end{widetext}

% \documentclass[12pt]{article}
%
%\onehalfspacing
%\thispagestyle{empty}

%===========< FIGURE 1 >=========================================
\begin{figure*}[t]
\includegraphics[width=2.0\columnwidth]{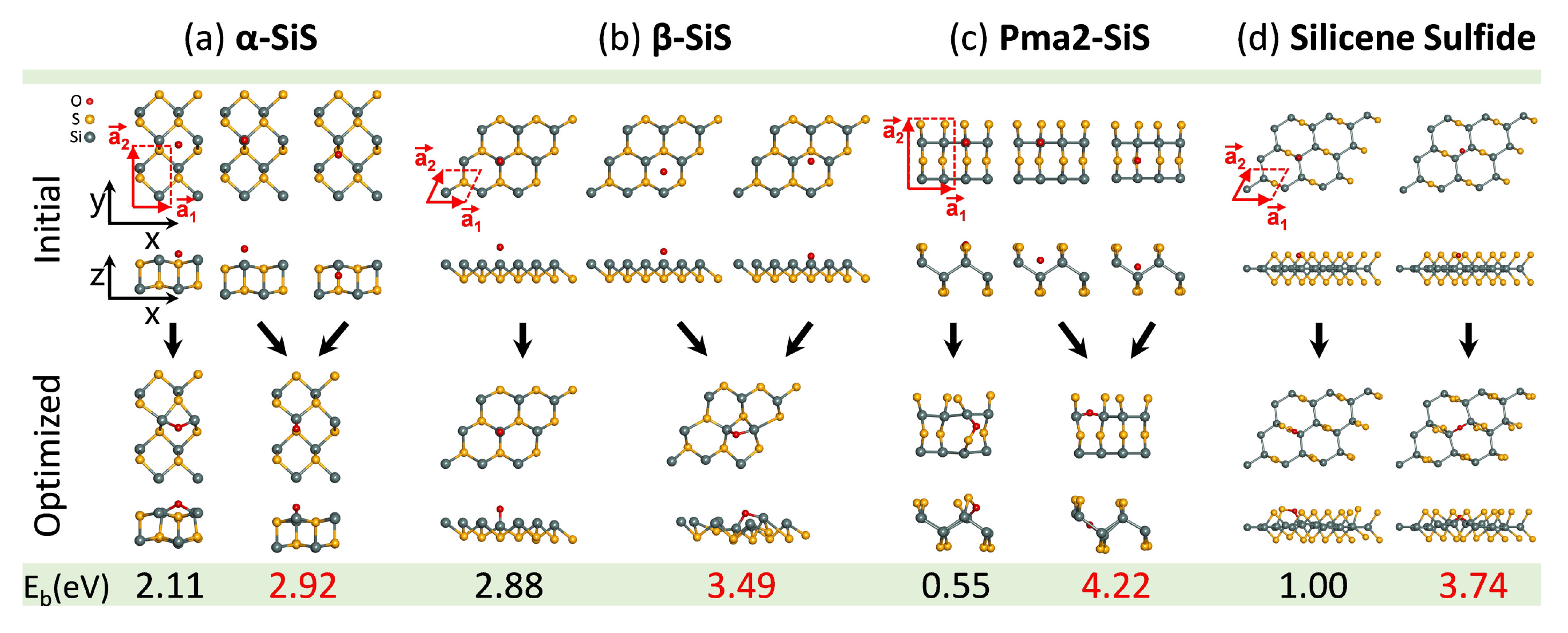}
\caption{Possible configurations of isolated oxygen atom absorbed on the monolayer of (a) $\alpha$-SiS, (b) $\beta$-SiS, (c) Pma2-SiS, (d) Silicene Sulfide. Ball-and-stick model in the top view and side view of initial structures are shown in the first row and that of optimized structures in the second row. Lattice vectors are indicated by red arrows in the top view of the initial structures. Calculated binding energies $E_b$ are given at the bottom and those for the most stable configurations are highlighted in red. %
\label{fig1} }
\end{figure*}
%===========< FIGURE 1 >=========================================

%===========< Table 1 >==========================================
\begin{table}[b]
\caption{Tab1: Calculated structural characters of pre-oxidized and post-oxidized SiS structures. $|\vec{a_1}|$ and $|\vec{a_2}|$ are the
in-plane lattice constants defined in Fig.~\protect\ref{fig1} and Fig.~\protect\ref{fig2}. $\angle\vec{a_1}\vec{a_2}$ is the angle between lattice vectors.}
\setlength{\tabcolsep}{4mm}
\begin{tabular}{lccr}
% \toprule
\hline %
& $|\vec{a_1}|$({\AA})        & $|\vec{a_2}|$({\AA}) %
& $\angle\vec{a_1}\vec{a_2}$ \\
\hline %
$\alpha$-SiS       & 3.31         & 4.38         & 90${^\circ}$ \\
$\beta$-SiS        & 3.30         & 3.30         & 60${^\circ}$ \\
Pma2-SiS           & 3.98         & 6.64         & 90${^\circ}$ \\
Silicene Sulfide   & 4.56         & 4.01         & 63.88${^\circ}$ \\
$\alpha$-SiSO      & 3.41         & 4.75         & 90${^\circ}$ \\
$\beta$-SiSO       & 2.87         & 3.33         & 64.47${^\circ}$ \\
Pma2-SiSO          & 5.23         & 6.63         & 90${^\circ}$ \\
Silicene Sulfide-O & 5.21         & 4.95         & 61.63${^\circ}$ \\ \hline %
\end{tabular}
\label{table1}
\end{table}
%===========< Table 1 >==========================================

\section*{I. Introduction}

There is a dramatically growing interest in two-dimensional (2D) materials since monolayer graphene was exfoliated successfully~\cite{Novoselov04S}. Phosphorene, a monolayer of black phosphorus, has attracted significant attention since 2014 due to its semiconducting character and high carrier mobility~\cite{Li2014,DT229,Koenig14,Avouris14}. As the isoelectric counterpart of phosphorene, monolayer silicon monosulfide (SiS) has been predicted shortly afterwards~\cite{DT247,yang16NL} and the group-IV monochalcogenide family (GeSe, SnSe, SnTe, ...) have become an important group of the 2D materials with their interesting novel properties~\cite{Gomes15prb,Kamal16prb,fei15apl,GeS15prb,SnSe18prm,Fei16prl,Chang16scie,
SnTe18prl,Salvador18prb,zhao14nature,Shafique17sr,Jiang18jps,Haleoot17prl,xu17prb,
Rangel17prl,Xue17jacs,zhao17jmca}. The potential applications of 2D group-IV monochalcogenides in low-dimensional piezoelectrics~\cite{fei15apl,GeS15prb}, ferroelectrics~\cite{Fei16prl,Chang16scie,SnTe18prl,Salvador18prb}, thermoelectrics~\cite{zhao14nature,Shafique17sr}, electrochemics~\cite{Jiang18jps} and optoelectronics~\cite{Haleoot17prl,xu17prb,Rangel17prl,Xue17jacs,zhao17jmca} have been well explored in both theory and experiment.

The chemical reactivity, which is a significant problem in phosphorene~\cite{Hersam14,water_bpox2015,Ziletti15prl}, could also be a challenge in the application of mono-layer group-IV monochalcogenides, as a result of their similarity in structure. Point defects in SnS, SnSe, GeS and GeSe monolayers have been studied theoretically and shown that these materials were less prone to oxidation than phosphorene~\cite{Gomes16prb}. However, 2D SiS, consisting of lighter elements, could be much easier to be oxidized due to the smaller atomic radii and stronger chemical bonding to oxygen. On the other hand, oxidation could be an effective way of chemical modification for 2D materials to tailor their physical and electrical properties, such as graphene oxides~\cite{Eda2010AM,Robinson08NL,Robinson08NL2} and phosphorene oxides~\cite{PO15Nanoscale,PO15prb}.

% Executive summary

In this study, to fill the missing information for the oxidation of SiS, isolated oxygen impurities and fully oxidized structures of four stable 2D SiS structures are investigated by {\em ab initio} density functional calculations. Binding energies of oxygen impurities in all the four 2D SiS structures are found larger than that in phosphorene, due to the lower electronegativity of Si atoms. The most stable configurations of isolated oxygen impurities for different 2D SiS structures are decided and the corresponding 2D structures with saturated oxidation (SiSO) are predicted. Among all the fully oxidized structures, $\alpha$-SiSO is demonstrated to be the most stable by phonon spectra calculations and molecular dynamics (MD) simulations. Electronic structure calculations indicate that $\alpha$-SiSO is semiconducting with a direct band gap of ${\approx}2.28$~eV, which can be effectively tuned by uniaxial in-layer strain. Finally, $\alpha$-SiSO$_x$ ($x<1$) structures with unsaturated oxidation is studied. We find that the value of band gap and thermodynamic stability depend sensitively on the saturation level of oxygen.

\section*{II. Method and Computational Details}

We utilized {\em ab initio} density functional theory (DFT) as
implemented in the \textsc{VASP} code~\cite{{VASP},{VASP2},{VASP3}} to obtain insight into the equilibrium structure, stability and electronic
properties of SiS and corresponding oxidized structures reported here. Periodic boundary conditions are used throughout the study, with 2D structures represented by a periodic array of slabs separated by a vacuum region in excess of $15$~{\AA}. The reciprocal space is sampled by a fine grid~\cite{Monkhorst-Pack76} of $8{\times}8{\times}1$~$k$-points in the Brillouin zone of the unit cell with 16 atoms or its equivalent in other supercells. We used projector-augmented-wave (PAW) pseudopotentials~\cite{PAWPseudo} and the Perdew-Burke-Ernzerhof (PBE)~\cite{PBE} exchange-correlation functionals. Electronic wave function is expanded on a plane-wave basis set with cutoff energy of 500~eV. A total energy difference between subsequent self-consistency iterations below $10^{-5}$~eV is used as the criterion for reaching self-consistency. All geometries have been optimized using the conjugate-gradient method~\cite{CGmethod}, until none of the residual Hellmann-Feynman forces exceeded $10^{-2}$~eV/{\AA}. The phonon calculations are carried out using
the supercell approach, as implemented in the PHONOPY code~\cite{phonopy}. In the MD simulations, supercells including more than 100 atoms are used and the temperature of $300$~K and $500$~K is kept for longer than 3 ps with a time step of 3 fs.
%The calculation of phonon spectrum is studied with density functional %perturbation theory (DFPT), FORCE\_CONSTANTS file is structured by %Phonopy from the Hessian matrix in vasprun.xml file, which can be %reprocess to get phonon spectrum. We use Microcanonical (NVE) ensemble %(SMASS=-3) to simulate molecular dynamics calculations for %$\alpha$-SiSO and two different semi-oxidation $\alpha$-SiSO$_{0.5}$ %in 300K, 500K and 1000K. The electronic occupation details of valence %band maximum (VBM) and the conduction band minimum (CBM) are expressed %with partial charge density calcualtions.

%===========< FIGURE 2 >=========================================
\begin{figure*}[t]
\includegraphics[width=1.7\columnwidth]{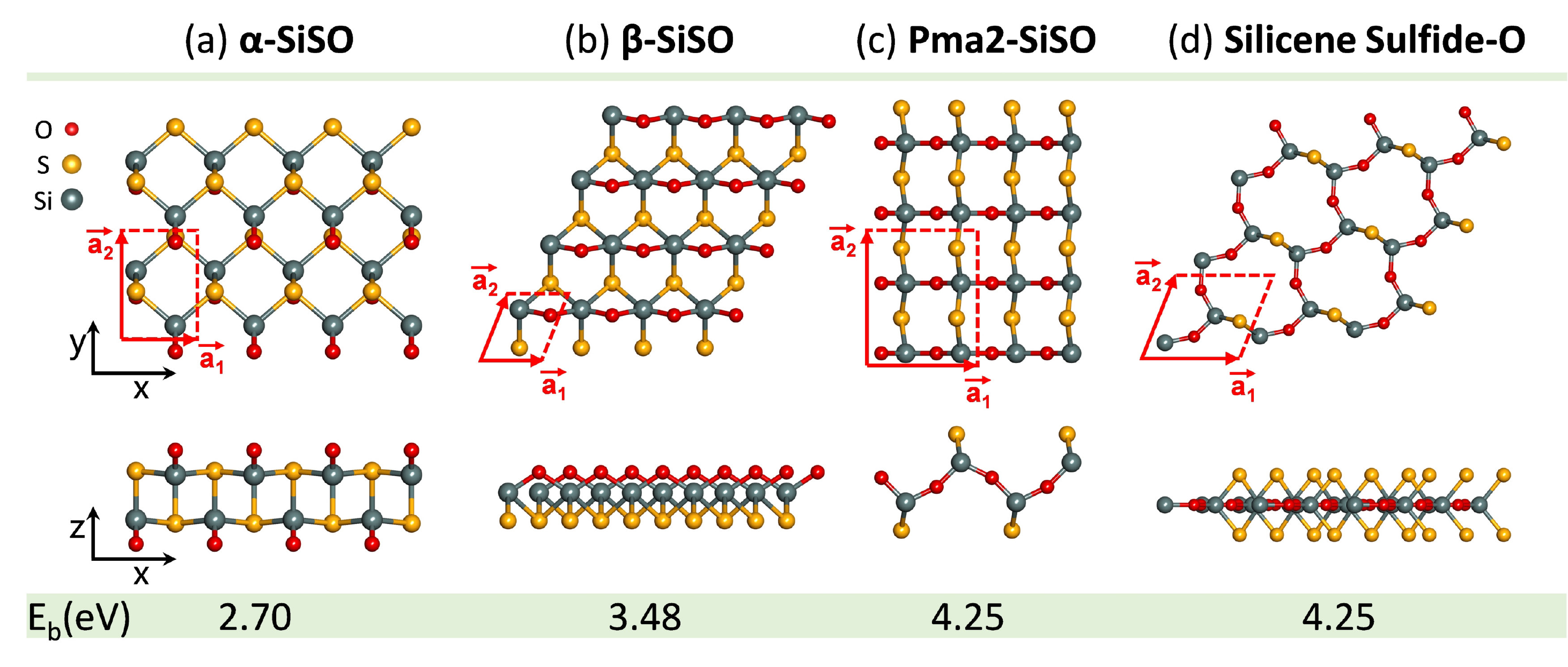}
\caption{Ball-and-stick model in the top view (1st row) and side view (2nd row) of equilibrium structures for monolayers of (a) $\alpha$-SiSO, (b) $\beta$-SiSO, (c) Pma2-SiSO and (d) Silicene sulfide-O. Lattice vectors are indicated by red arrows in the top view. Calculated binding energies $E_b$ are given at the bottom. %
\label{fig2} }
\end{figure*}
%===========< FIGURE 2 >=========================================

%===========< FIGURE 3 >=========================================
\begin{figure*}[t]
\includegraphics[width=1.8\columnwidth]{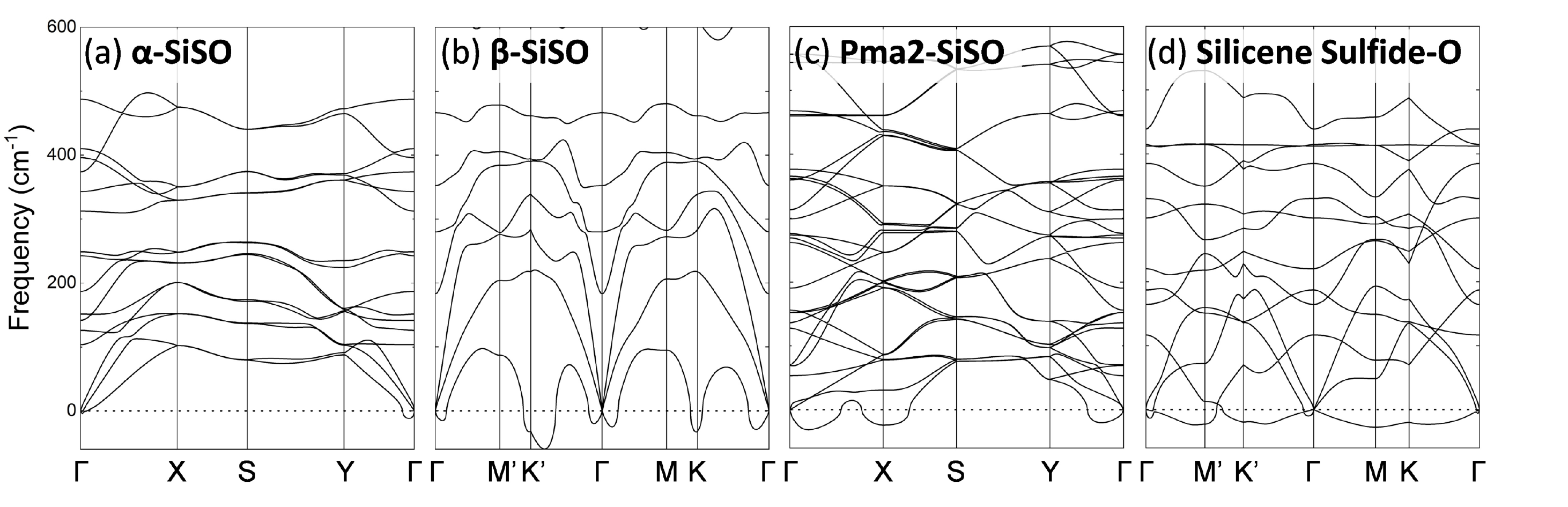}
\caption{Vibrational phonon spectra for monolayers of (a) $\alpha$-SiSO, (b)$\beta$-SiSO, (c)Pma2-SiSO and (d) Silicene Sulfide-O.%
\label{fig3} }
\end{figure*}
%===========< FIGURE 3 >=========================================

%===========< FIGURE 4 >=========================================
\begin{figure}[t]
\includegraphics[width=1.0\columnwidth]{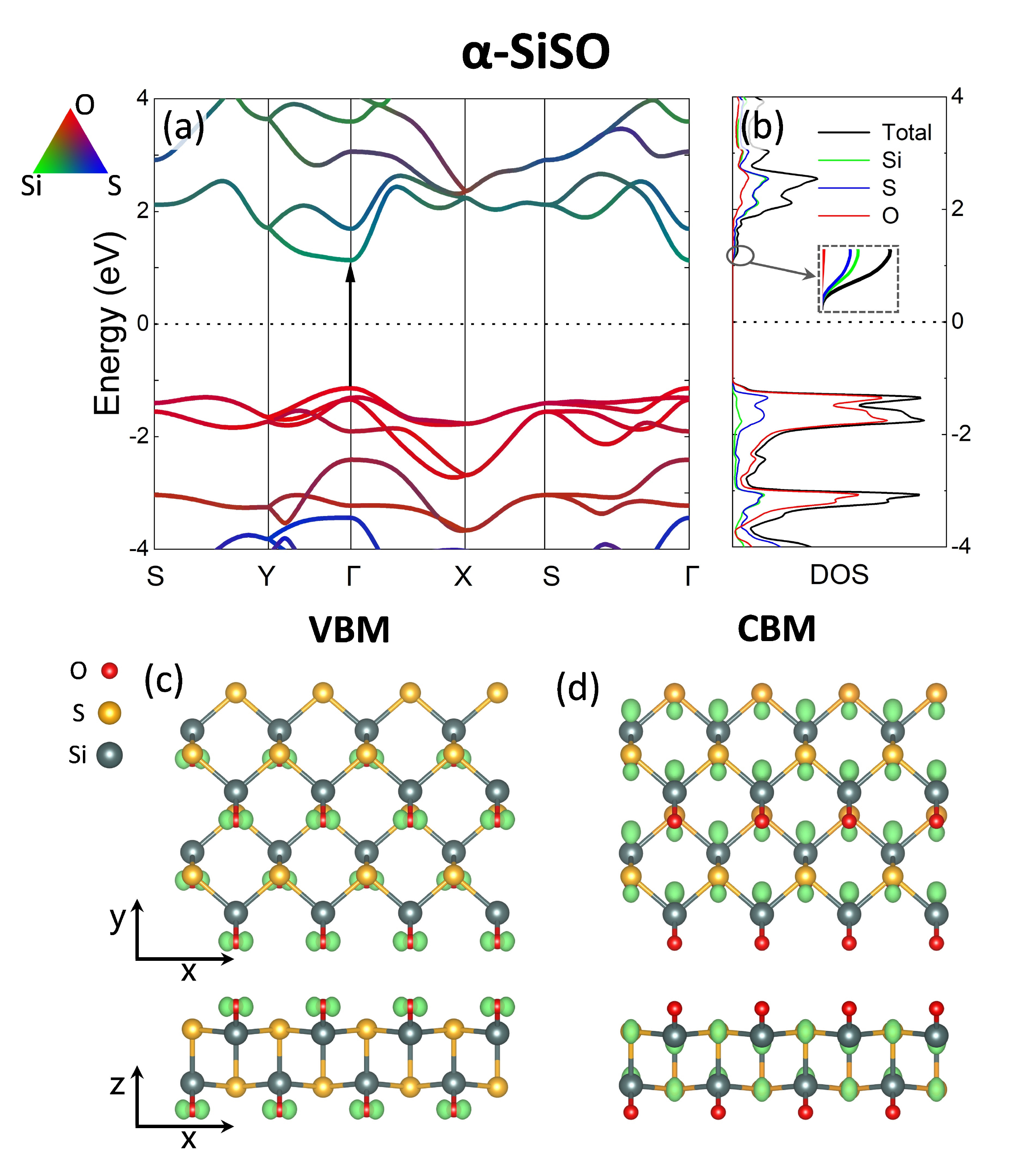}
\caption{(a) Electronic band structure, (b) projected density of states (PDOS), partial charge density associated with frontier states in the (c) valence band and (d) conduction band of $\alpha$-SiSO monolayer. Contributions of electrons from different atoms in (a), (b) are distinguished by color. A direct band gap is shown by a black arrow in (a). A figure zoomed in near the conduction band minimum (CBM) is shown as inset in (b). %
\label{fig4} }
\end{figure}
%===========< FIGURE 4 >=========================================

%===========< FIGURE 5 >=========================================
\begin{figure}[t]
\includegraphics[width=1.0\columnwidth]{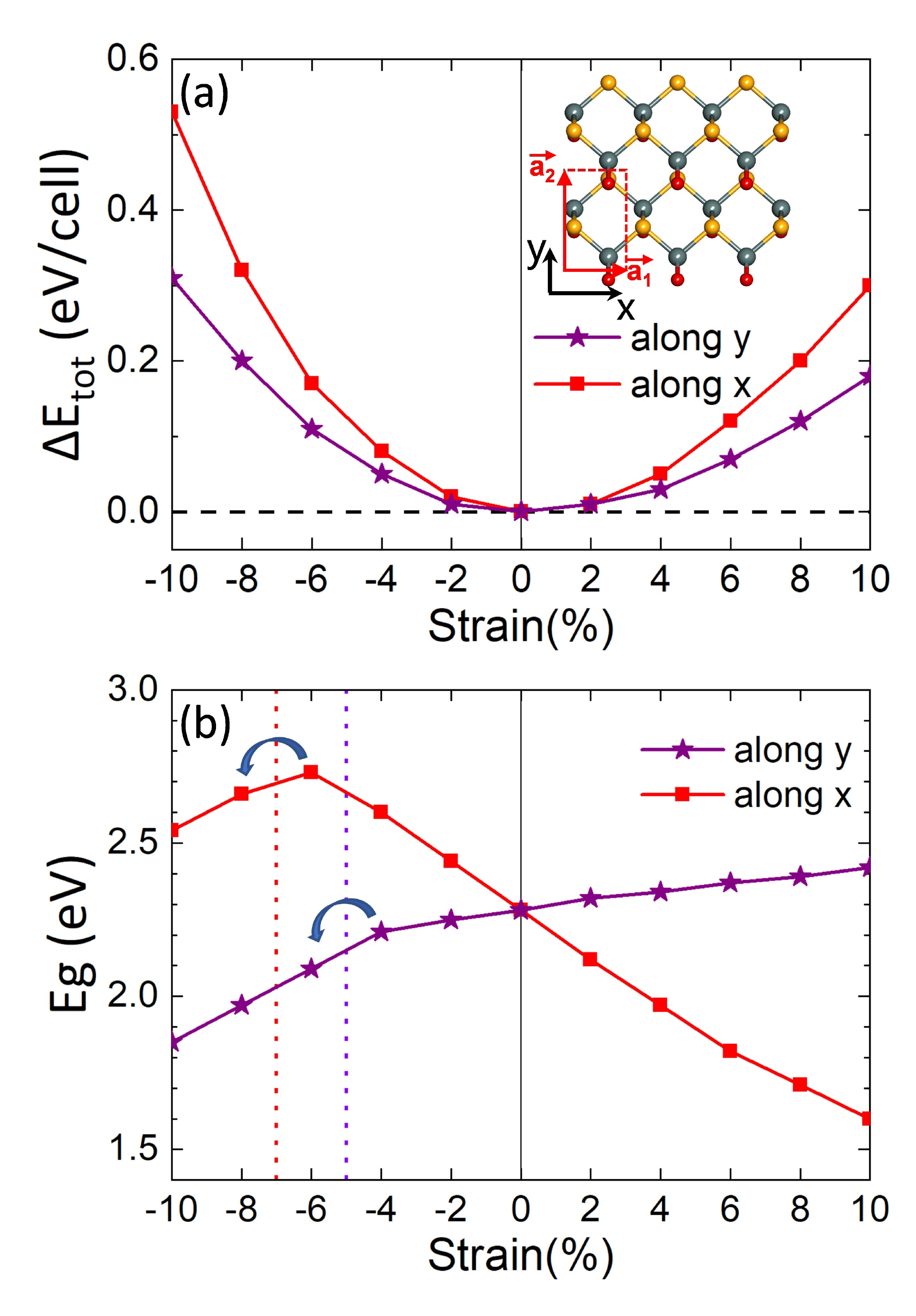}
\caption{Effect of uniaxial in-layer strain on (a)the relative total energy ${\Delta}E_{tot}$ and (b) the fundamental band gap $E_g$ in $\alpha$-SiSO monolayer. The strain directions defined in the inset of (a) are distinguished by color and symbols. Dashed vertical lines in (b) indicate a direct-to-indirect band gap transition. %
\label{fig5} }
\end{figure}
%===========< FIGURE 5 >=========================================

%===========< FIGURE 6 >=========================================
\begin{figure}[t]
\includegraphics[width=1.0\columnwidth]{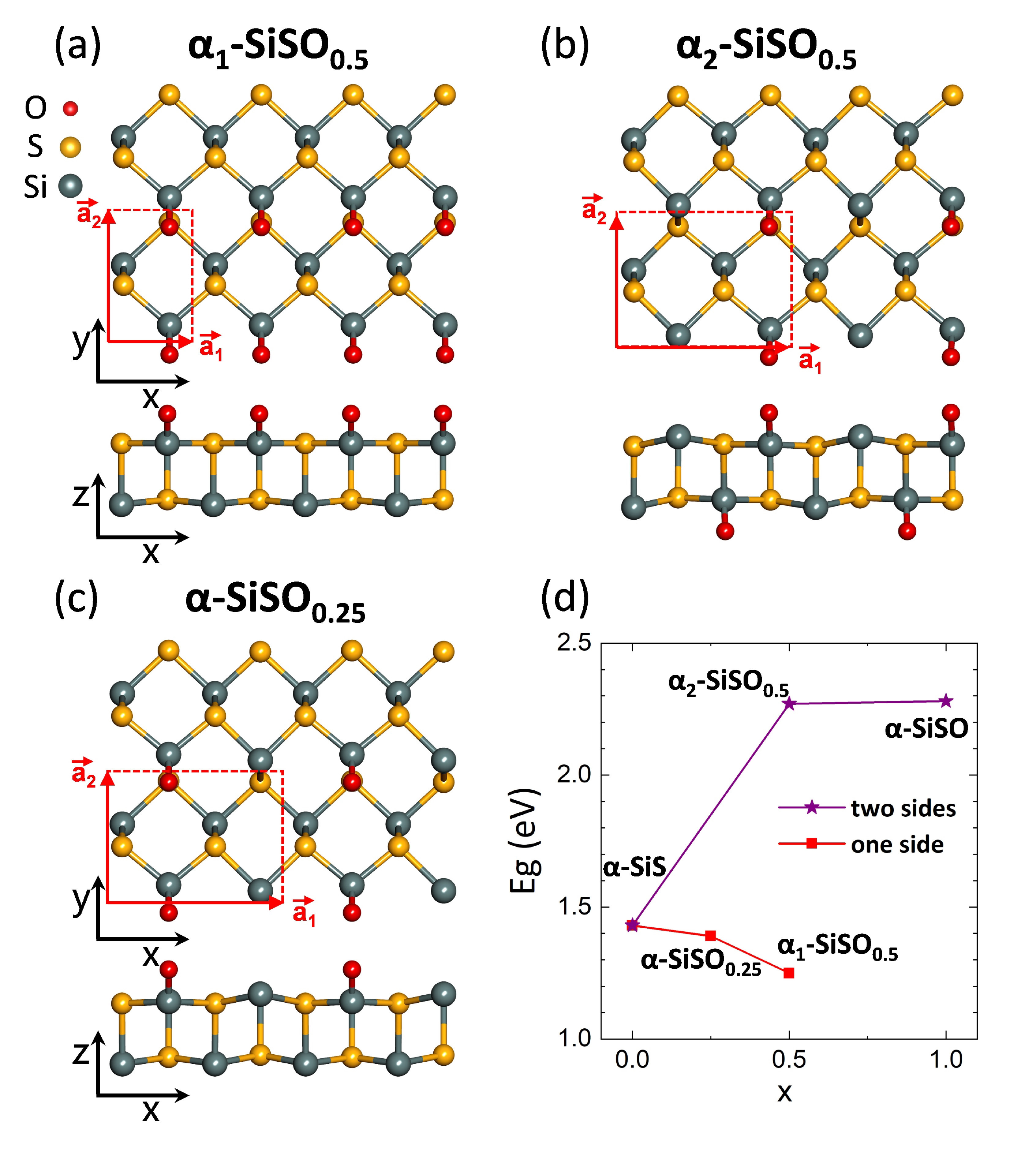}
\caption{Ball-and-stick model in the top view and side view of equilibrium structures for (a) $\alpha_1$-SiSO$_{0.5}$, (b) $\alpha_2$-SiSO$_{0.5}$ and (c) $\alpha$-SiSO$_{0.25}$. Lattice vectors are indicated by red arrows in the top view. (d) The value of band gap $E_g$ as a function of $x$ for $\alpha$-SiSO$_x$ monolayers. Structures with oxidation on one single side and both sides are distinguished by color and symbols. %
\label{fig6} }
\end{figure}
%===========< FIGURE 6 >=========================================

%===========< FIGURE 7 >=========================================
\begin{figure}[t]
\includegraphics[width=1.0\columnwidth]{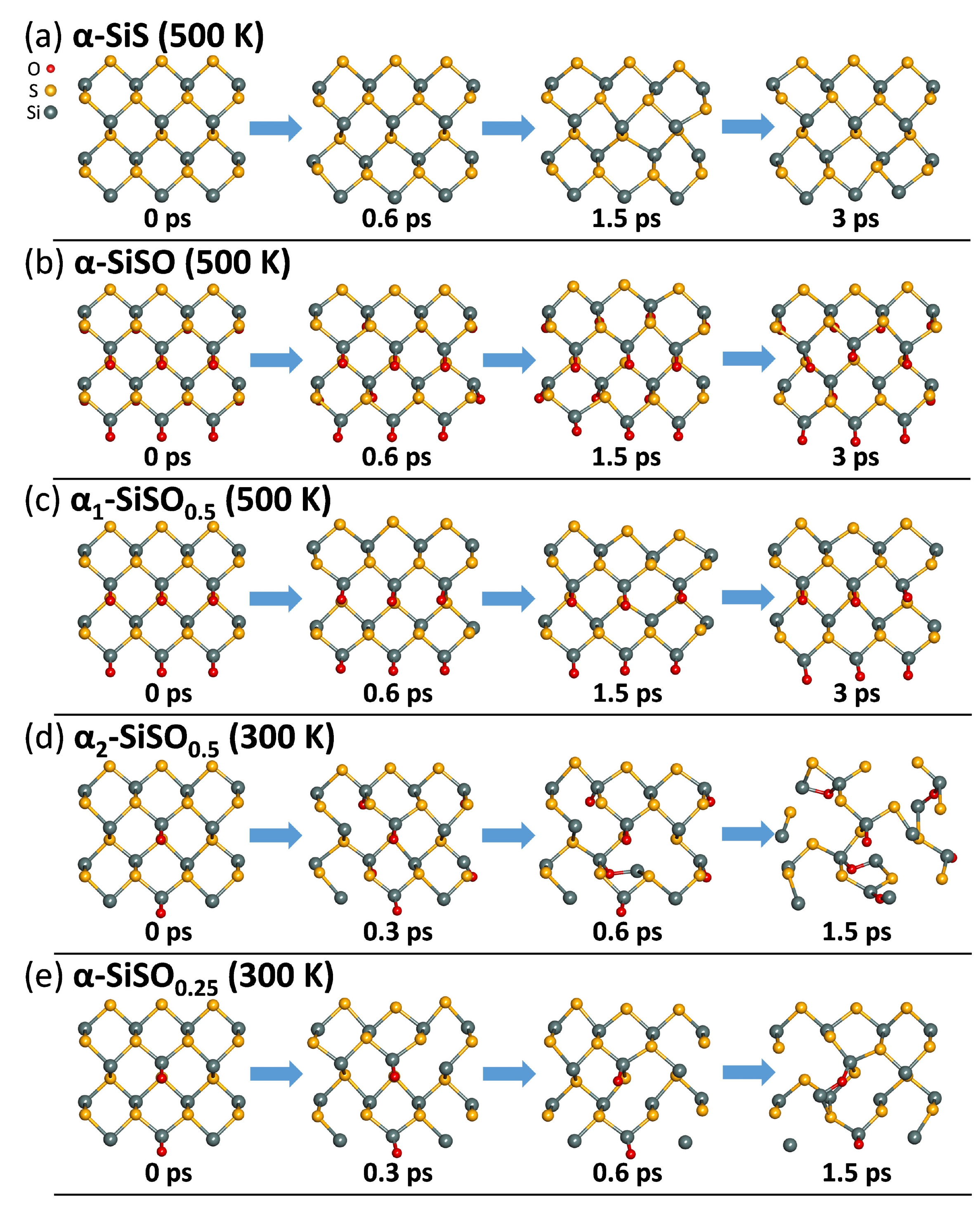}
\caption{Snap shots at different time steps of canonical molecular dynamics (MD) simulations depicting structural changes in (a) $\alpha$-SiS, (b) $\alpha$-SiSO, (c) $\alpha_1$-SiSO$_{0.5}$, (d) $\alpha_2$-SiSO$_{0.5}$ and (e) $\alpha$-SiSO$_{0.25}$ monolayers under the temperature of 300~K. %
\label{fig7} }
\end{figure}
%===========< FIGURE 7 >=========================================

\section*{III. Results and Discussions}

\subsection*{A. Isolated cxygen impurities}

Four different 2D SiS structures predicted in previous works were considered, i.e. $\alpha$-SiS, $\beta$-SiS~\cite{DT247} and Pma2-SiS, silicene sulfide~\cite{yang16NL}. As the optimized structures for the monolayers shown in the first row of Fig.~\ref{fig1}, $\alpha$-SiS and $\beta$-SiS are isoelectronic with black and blue phosphorene~\cite{DT230,DT232}, respectively. Therefore they are similar to phosphorene in geometry with all atoms threefold coordinated. However, in Pma2-SiS and silicene sulfide all atoms obey ``octet rule'' with Si fourfold coordinated and S twofold coordinated. $\alpha$-SiS and Pma2-SiS have orthorhombic lattices, $\beta$-SiS has a honeycomb lattice and silicene sulfide has a stretched honeycomb lattice with the angle of $63.88{^\circ}$ between lattice vectors. All the calculated structural characteristics are summarized in Table~\ref{table1} and consistent with previous works~\cite{DT247,yang16NL}.

To investigate the resistance to oxidation, interactions between isolated oxygen atom with all the selected SiS structures were calculated. As shown in Fig.~\ref{fig1}(a), initially three possible configurations for O atom approaching $\alpha$-SiS are considered: the ring site which is above the center of the hexagonal ring, the Si-top site which is on top of the Si atom, the interstitial site which is close to the Si-S bond and interred into the layer. Interestingly, both the Si-top and interstitial sites prefer the dangling site attaching to Si atoms in the optimized configurations. The initially ring site was optimized to a Si-O-Si bridge site. The binding energy $E_b$ per oxygen atom, which is defined as: $E_b = E_{SiS} + E_{O_2}/2 – E_{SiS+O}$, is $2.92$~eV for the Si-dangling site and $2.11$~eV for the Si-O-Si bridge site in the equilibrium geometry.

Similarly, possible configurations of oxygen impurities in the other three structures are shown in Fig.~\ref{fig1}(b)-(d). In $\beta$-SiS, initially the Si-top site, ring site and Si-O-Si bridge site configurations were considered and the equilibrium configurations of Si-top site with $E_b$ equalling $2.88$~eV and Si-O-Si bridge site with $E_b$ equalling $3.49$~eV were obtained. In Pma2-SiS, initial configurations with O atom embedded in the hollows of S-Si-S, Si-Si-Si and Si-S-Si were optimized and equilibrium configurations of Si-O-S bridge site with $E_b$ equalling $0.55$~eV and Si-O-Si bridge site with $E_b$ equalling $4.22$~eV were obtained. In silicene sulfide, initial configurations of Si-top site and Si-O-Si bridge site were optimized and equilibrium configurations of Si-O-S bridge site with $E_b$ equalling $1.00$~eV and Si-O-Si bridge site with $E_b$ equalling $3.74$~eV were obtained. From all these results we found that (i) in all the structures the oxygen atom prefer Si atoms to S atoms due to the lower electronegativity of Si, (ii) the most stable configuration of oxygen impurity is Si-dangling site in $\alpha$-SiS and is Si-O-Si bridge site in the rest, and (iii) bind energies of the most stable oxygen impurities for all the 2D SiS structures are much larger than that for other 2D group-IV monochalcogenides, such as SnS ($0.88$~eV), SnSe ($0.75$~eV), GeS ($1.47$~eV), GeSe ($1.22$~eV)~\cite{Gomes16prb} and also larger than that for phosphorene ($2.08$~eV)~\cite{Ziletti15prl}. The large binding energy indicates the inclination to oxidize in the air for the selected 2D SiS structures.

\subsection*{B. Fully oxidized structures}

Here we chose the most stable configuration of oxygen impurity in each SiS structure and further studied their corresponding SiSO structures with saturated oxidation. The optimized geometry of $\alpha$-SiSO monolayer is shown in Fig.~\ref{fig2}(a). Each Si atom is attached by one O atom and the bind energy $E_b$ per oxygen atom equalling $2.70$~eV is obtained, which is a little smaller than the value of $2.92$~eV for isolated oxygen impurity. This is because of the repulsive interaction between neighboring O atoms in the fully oxidized structure, with details discussed in the Supplemental Material~\cite{SM-SiS18}. In $\beta$-SiSO, neighboring Si atoms are connected by O atoms with $E_b$ equalling $3.48$~eV, which is almost the same as isolated bridge-site oxygen impurity, as shown in Fig.~\ref{fig2}(b). Pma2-SiSO and silicene sulfide-O structures are similar, where all the pristine Si-Si bonds are broken and new Si-O-Si bonds are formed, instead. Both of them got larger values of $E_b$ (4.25~eV) than those for isolated oxygen impurities due to the vanishing of distortion in the uniformly oxidized structures. Structural characteristics for all the SiSO structures are summarized in Table~\ref{table1}. We found that after oxidation the unit cell of $\alpha$-SiSO become larger than before. The main reason for this is that the extra Si-O bonds modify the hybridization of orbitals in Si, enlarge the angles between the three Si-S bonds around each Si atom and finally extend the area of the unit cell in $x-y$ plane. The unit cell of $\beta$-SiSO shrinks in $x$-direction due to the extra Si-O-Si bonding. Both Pma2-SiSO and silicene sulfide-O get larger unit cells as a result of O atoms embedding into the lattices.

As a robust way to illustrate the structural stability, vibrational phonon spectra calculations were carried out for the four equilibrium SiSO structures shown in Fig.~\ref{fig2}. The results of phonon dispersion for the monolayers are shown in Fig.~\ref{fig3}. We found that dynamically the most stable structure is $\alpha$-SiSO, with essentially no imaginary modes. The tiny ``U'' shape feature near $\Gamma$-Point in the out-of-plane acoustic (ZA) mode is caused by the artifact of the method but has nothing to do with a structural instability, which is quite common in other 2D systems~\cite{DT255,Yu16jmcc}. In contrast, significant negative frequencies not only near $\Gamma$-Point were found in the phonon dispersion of $\beta$-SiSO, Pma2-SiSO and silicene sulfide-O, which indicates the dynamic instability of freestanding monolayers. 2D $\alpha$-SiSO structure with the best dynamic stability as shown here will be focused on in the following discussions.
%little negative frequencies at $\Gamma$, not more than -10cm$^{-1}$, %which can be considered stable approximately. $\beta$-SiSO
%Fig.~\ref{fig3}(b) present -15cm$^{-1}$ negative frequencies at
%$\Gamma$, Pma2-SiSO (Fig. 3(c)) and Silicene Sulfide-O
%Fig.~\ref{fig3}(d) perform obvious negative frequencies bigger than
%-27cm$^{-1}$ at whole reciprocal space.

% \subsection*{Electronic properties}

As a previously unexplored 2D material, the electronic properties of $\alpha$-SiSO were investigated. The electronic band structure and the associated projected density of states (PDOS) are shown in Fig.~\ref{fig4}(a) and (b). Our DFT-PBE calculations show that $\alpha$-SiSO monolayer is a semiconductor with a direct band gap of 2.28~eV. Both the valence band maximum (VBM) and the conduction band minimum (CBM) are located at $\Gamma$-Point. Contributions of electrons from different atoms are indicated by different colors in the band structure and PDOS plots. PDOS near the CBM is easier to see from the amplified figure in the inset of Fig.~\ref{fig4}(b). We found that the VBM is dominated by electrons from O, while the CBM is dominated by electrons from Si and S. More details about the electron characters of the VBM and CBM were explored by partial charge density plotting associated with frontier states, as depicted in Fig.~\ref{fig4}(c) and (d). The charge distribution in the VBM was found surrounding O atoms and appears a dumbbell-like shape along $x$-direction, which is a signature of the domination of $p_x$ orbitals from oxygen. The charge distribution in the CBM is much different and appears a polarized $s$ orbitals along $y$-direction surrounding Si and S atoms, which is the signature of a hybridization of $s$ and $p_y$ orbitals from Si and S. The results can be further verified by detailed PDOS plotting for different orbitals from each single element shown in Supplemental Material~\cite{SM-SiS18}. We also found from Fig.~\ref{fig4}(a) that the band dispersion is much flatter along ${\Gamma}-Y$ direction than that along ${\Gamma}-X$ direction near the CBM at $\Gamma$-Point, which indicates a significant anisotropic carrier effective mass. This can be understood because the frontier states at the CBM from Si and S are very close to each other and can have a strong interaction along $x$-direction, as shown in Fig.~\ref{fig4}(d).

Similar to other non-planar 2D systems like phosphorene and SiS, $\alpha$-SiSO is susceptible to even minute in-plane strain. To quantify the effect, the overall dependence of the stability and the fundamental band gap on the tensile and compressive in-layer strain of $\alpha$-SiSO is calculated and presented in Fig.~\ref{fig5}. We have considered uniaxial strain along $x$- and $y$-direction, as defined in the inset of Fig.~\ref{fig5}(a). A distinct anisotropy of the strain energy with respect to the strain direction is shown in Fig.~\ref{fig5}(a), which results from the distinct structural anisotropy. Similar to other phosphorene-like buckled structures, the system appears softer when strained along the armchair-like direction ($y$-direction) than the other one. We found that the overall strain energy ${\Delta}E_{tot}$ is quite small, which is about 0.53~eV/cell or 0.30~eV/cell under 10\% compressive or tensile strain, respectively, along $x$-direction. The value for strain along the softer $y$-direction is almost halved, which is about 0.30~eV/cell or 0.18~eV/cell under 10\% compression or stretch, respectively.

The effect of uniaxial in-layer strain on the fundamental band gap obtained by our DFT-PBE calculations is shown in Fig.~\ref{fig5}(b). We found that the band gap reduces down to 1.60~eV under the ${\lesssim}$10\% stretch along $x$-direction. When compressed, the band gap increases and reach the largest value of 2.73~eV under 6\% compression. When the compressive strain exceeds 6\%, the value of band gap will decrease and in the meantime the direct gap is transformed to an indirect gap. A generally opposite effect was found along $y$-direction with the value of the band gap varying between 1.85~eV under 10\% compressive strain and 2.42~eV under 10\% tensile strain. A direct-to-indirect gap transition was also found by a ${\gtrsim}4$\% compression along $y$-direction.

\subsection*{C. Effects of oxidation saturation}

Effects of the oxidation degree on 2D $\alpha$-SiSO were examined by the calculations for $\alpha$-SiSO$_x$ ($0<x<1$) structures. Monolayers of $\alpha_1$-SiSO$_{0.5}$, $\alpha_2$-SiSO$_{0.5}$ and $\alpha$-SiSO$_{0.25}$ were optimized and the equilibrium geometries are depicted in Fig.~\ref{fig6}(a)-(c). In $\alpha_1$-SiSO$_{0.5}$ O atoms are attached to Si atoms on only one single side of $\alpha$-SiS while in $\alpha_2$-SiSO$_{0.5}$ O atoms are attached on both sides with the density halved. Finally, in $\alpha$-SiSO$_{0.25}$ O atoms are only attached to half of the Si atoms on one single side. Our DFT-PBE results for electronic band gap of selected $\alpha$-SiSO$_x$ ($0{\le}x{\le}1$) monolayers are summarized in Fig.~\ref{fig6}(d). Interestingly, we found that the value of the band gap increases with the value of $x$ if oxidation happens on both sides of the monolayer (from 1.43~eV for $\alpha$-SiS to 2.28~eV for $\alpha$-SiSO). Contrarily, if oxidation happens on a single side, the value of band gap decreases when $x$ increases (from 1.43~eV for $\alpha$-SiS to 1.25~eV for $\alpha_1$-SiSO$_{0.5}$). The detailed band structure and PDOS plots for $\alpha$-SiSO$_x$ ($0<x<1$) are shown in Supplemental Material~\cite{SM-SiS18}. This finding provides another way to effectively tune the electronic properties of the pristine structure by appropriate oxidation.

Besides the electronic properties, the effect of oxygen saturation on the thermodynamic stability for 2D $\alpha$-SiSO$_x$ ($0{\le}x{\le}1$) structures were investigated by MD simulations. Our canonical MD simulations for monolayers of $\alpha$-SiS, $\alpha$-SiSO, $\alpha_1$-SiSO$_{0.5}$, $\alpha_2$-SiSO$_{0.5}$ and $\alpha$-SiSO$_{0.25}$ were performed under the temperature 300~K and 500~K, respectively. We found that $\alpha$-SiS, $\alpha$-SiSO and $\alpha_1$-SiSO$_{0.5}$ are stable under both $300$~K and $500$~K. As the snapshots during the MD process under $500$~K shown in Fig.~\ref{fig7}(a)-(c), all the three structures keep their geometries nearly unchanged during a simulation time up to 3~ps, which indicates good thermodynamic stabilities at high temperature. However, the other two structures are not stable under either $300$~K or $500$~K. The snapshots during the MD process under $300$~K for $\alpha_2$-SiSO$_{0.5}$ and $\alpha$-SiSO$_{0.25}$ are shown in Fig.~\ref{fig7}(d) and (e). We can see that $\alpha_2$-SiSO$_{0.5}$ gets significantly distorted and finally degraded after 0.6~ps. The reason for this is that in $\alpha_2$-SiSO$_{0.5}$, each oxidized surface is not saturated. The O atom attached to one Si is easily attracted by the neighboring exposed Si atoms and forms a bridge configuration. This will cause a significant distortion of the global structure and finally break the structure. The situation is much different in $\alpha$-SiSO and $\alpha_1$-SiSO$_{0.5}$ because it is saturated on each oxidized surface. Each O atom is restricted bonding to one Si and cannot move away due to the repulsive interaction from the neighboring O atoms, which was mentioned above and also in Supplemental Material~\cite{SM-SiS18}. In $\alpha$-SiSO$_{0.25}$, Si-O-Si bridge configurations are also found after 0.6~ps. The structure is less distorted than the case of $\alpha_2$-SiSO$_{0.5}$ due to the smaller degree of oxidation saturation, but still on the brink of degradation. The fully oxidized surface of $\alpha$-SiS with good stability under high temperature indicates the possibility of application for $\alpha$-SiSO as the protecting layer in experimental preparation of the pristine 2D SiS structure.

\section*{IV. Summary and Conclusions}

In summary, we have investigated isolated oxygen impurities of 2D $\alpha$-SiS, $\beta$-SiS, Pma2-SiS and silicene sulfide structures by {\em ab initio} DFT calculations. Binding energies of oxygen impurities for all the four 2D SiS structures are found larger than those for phosphorene, due to the lower electronegativity of Si atoms. The most stable configurations of isolated oxygen impurities for the four different 2D SiS structures are decided by the binding energy of oxygen and the corresponding 2D SiSO structures with saturated oxidation are predicted. Among all the fully oxidized structures, $\alpha$-SiSO is demonstrated to be the most stable by phonon spectra calculations. Electronic structure calculations indicate that $\alpha$-SiSO is semiconducting with a direct band gap of ${\approx}2.28$~eV. The value of band gap can be effectively tuned by uniaxial in-layer strain and a direct-to-indirect gap transition can happen under compressive strain along both directions. The value of band gap for 2D $\alpha$-SiSO$_x$ ($0{\le}x{\le}1$) structures was found dramatically increasing with the value of $x$ when oxidized on both sides while decreasing when oxidized on a single side. Our MD simulations show that $\alpha$-SiSO and $\alpha_1$-SiSO$_{0.5}$ structures with saturated oxidation on one or both surfaces are as stable as $\alpha$-SiS under high temperature at $500$~K. While $\alpha_2$-SiSO$_{0.5}$ and $\alpha$-SiSO$_{0.25}$ structures with unsaturated oxidation surfaces  have poorer thermodynamic stability and are easy to get degraded even at 300~K, due to the attractive interaction between O atoms and the neighboring exposed Si atoms. The fully oxidized surface of $\alpha$-SiS with good stability at high temperature could be used as the protecting layer for the pristine 2D structure in application.

\begin{acknowledgments}
This study was supported by National Natural Science Foundation of China (NSFC) under Grant NO. 61704110, 11674055, 11834002, by Fundamental Research Funds for the Central Universities and by Shuangchuang Doctor Program of Jiangsu Province.
\end{acknowledgments}

% \bibliographystyle{apsrev4-1}
% \bibliography{SiSO18}
% \end{document}

\end{document}